# Fully-staggered-array bulk Re-Ba-Cu-O short-period undulator: large-scale 3D electromagnetic modelling and design optimization using *A-V* and *H*-formulation methods

Kai Zhang[1], Mark Ainslie[2], Marco Calvi[1], Ryota Kinjo[3], Thomas Schmidt[1]

[1] Insertion Device Group, Photon Science Division, Paul Scherrer Institute, Villigen, 5232, Switzerland

[2] Bulk Superconductivity Group, Department of Engineering, University of Cambridge, Cambridge, CB2 1PZ, United Kingdom

[3] Advanced X-Ray Laser Group, RIKEN SPring-8 Center, Hyogo, 679-5148, Japan

Email: kai.zhang@psi.ch; mda36@cam.ac.uk

## Abstract

The development of a new hard *x*-ray beamline I-TOMCAT equipped with a 1-meter-long short-period bulk high-temperature superconductor undulator (BHTSU) has been scheduled for the upgrade of the Swiss Light Source (SLS 2.0) at the Paul Scherrer Institute (PSI). The very hard x-ray source generated by the BHTSU will increase the brilliance at the beamline by over one order of magnitude in comparison to other state-of-the-art undulator technologies and allow experiments to be carried out with photon energies in excess of 60 keV. One of the key challenges for designing a 1-meter-long (100 periods) BHTSU is the large-scale simulation of the magnetization currents inside 200 staggered-array bulk superconductors. A feasible approach to simplify the electromagnetic model is to retain five periods from both ends of the 1-meter-long BHTSU, reducing the number of degrees of freedom (DOFs) to the scale of millions. In this paper, the theory of the recently-proposed 2D *A-V* formulation-based backward computation method is extended to calculate the critical state magnetization currents in the ten-period staggered-array BHTSU in 3D. The simulation results of the magnetization currents and the associated undulator field along the electron beam axis are compared with the well-known 3D *H*-formulation and the highly efficient 3D *H-φ* formulation method, all methods showing excellent agreement with each other as well as with experimental results. The mixed *H-φ* formulation avoids computing the eddy currents in the air subdomain and is significantly faster than the full *H*-formulation method, but is slower in comparison to the *A-V* formulation-based backward computation. Finally, the fastest and the most efficient *A-V* formulation in ANSYS 2020R1 Academic is adopted to optimize the integrals of the undulator field along the electron beam axis by optimizing the sizes of the end bulks.

Keywords: HTS modelling, backward computation, critical state model, finite element method, H-formulation, bulk superconductors, undulator

___

## 1. Introduction

In 2004, Tanaka et al. first proposed the concept of the bulk high-temperature superconductor undulator (BHTSU) in which a dipole field was utilized to magnetize in-situ a series of rectangular high-temperature superconductors (HTS) [1]. In 2008, Kinjo et al. proposed the concept of the staggered-array BHTSU by using a superconducting solenoid to magnetize a series of staggered-array superconducting half-moon-shaped disks [2], and in 2013, they demonstrated for the first time an undulator field $B_0$ of 0.85 T in a 10-mm period, 4-mm gap BHTSU prototype subjected to an external field change $\Delta B$ of 4 T [3]. In 2019, Calvi et al. demonstrated an undulator field $B_0$ of ~0.85 T in a 10-mm period, 6-mm gap BHTSU prototype subjected to an external field change $\Delta B$ of 7 T [4]. Theoretically, the value of $B_0$ could be doubled for a 4-mm gap. Based on numerical calculations, an undulator field $B_0$ above 2 T can be obtained at 10-mm period, 4-mm gap with an external field change $\Delta B$ of 10 T, assuming no quenches occur during the field-cooled (FC) magnetization process. In late 2020, the Paul Scherrer Institute (PSI) scheduled the development of a 1-m long BHTSU with a 10-mm period and 4-mm gap for installation in the new I-TOMCAT microscopy tomography beamline planned for the upgraded Swiss Light Source (SLS 2.0) [5]. The x-ray source generated by the BHTSU will increase the brilliance of the beamline by well over one order of magnitude in comparison to state-of-the-art cryogenic permanent magnet undulator (CPMU) technology [6]. The new I-TOMCAT beamline would allow experiments to be carried out for very hard x-rays with photon energies in excess of 60 keV, thereby extending the photon range by almost a factor of two with respect to comparable instruments in medium-energy synchrotrons [7].

One of the key challenges for designing a 1-m long (100 periods) staggered-array BHTSU is the large-scale 3D modelling of the magnetization currents inside the 200 bulk superconductors comprising it. Considering the central superconducting bulks can generate a uniform sinusoidal undulator field, a simplified way to optimally design the 1-m long BHTSU is to retain five periods from both ends. The problem then becomes the optimization of the field integrals along the electron beam. In the case of using the finite-element method (FEM), a fine meshing of the superconducting bulks and the surrounding air subdomain is necessary to obtain smooth and accurate computational results for the undulator field. This, unfortunately, results in millions of elements, and thus degrees of freedom (DOFs). State-of-the-art FEM models implementing the ***E-J*** power law to simulate the superconductor's nonlinear resistivity can solve such complex 3D HTS problems, but can take a significant amount of computation time [8][9][10][11]. Other available numerical methods for critical state modelling are reviewed in several references [9][12][13][14], but in the following, the current state-of-the-art in numerical methods for modelling the critical state in bulk superconductors with respect to large-scale 3D HTS modelling is summarized.

In 1994, Bossavit first proposed the variational formulation method for the generalized Bean model. The electric field *E* was treated as a subdifferential of a critical energy density; the value of *E* was set to either zero if the current density |*J*| was lower than the critical current density $J_c$ or infinity if |*J*| was larger than $J_c$. This method was further developed by Prigozhin et al. [15][16][17][18][19], Elliott et al. [20] and Barrett et al. [21] to numerically analyze the critical state in type-II superconductors in 2D. In 2017, Pardo et al. further developed the variational method for computing critical state currents in a 3D cubic bulk superconductor, naming it the Minimum Electro-Magnetic Entropy Production (MEMEP) method [14]. The magnetic energy minimization (MEM) method, similar to the variational formulation, was first discussed by Badia et al. in 1998, and then adopted by Sanchez to compute the magnetic properties of finite superconducting cylinders and by Pardo et al. to study the magnetic properties of superconducting strips [22][23][24][25]. This approach employed a number of iteration steps to find the maximum penetration elements, therefore, minimizing the magnetic energy in the superconductor. In 1997, Nagashima et al. adopted a sand-pile model, which was proposed by Tamegai et al. in 1993, for the numerical calculation of the magnetic field distribution in Y-Ba-Cu-O bulks [26][27]. This analytical approach assumes the magnetization current flows along the periphery, namely in square current loops in rectangular conductors and circular current loops in cylindrical conductors [28][29][30]. In 1995, Brandt proposed the integral method based on an equation of motion for the current density and an ***E-J*** power law for computing the critical state in bulk superconductors with rectangular cross-section [31]. This integral method was extended by Brandt in 1996 to compute more realistic cases and by Bouzo et al. to solve 3D magnetization problems [32][33]. In 1976, Witzeling proposed the circuit method to compute the screening currents inside a superconducting cylinder, assuming the superconductors as an array of parallel wires and creating a relation between different current loops based on the law of induction [34]. In fact, this was the very first numerical method which solved the critical state currents in superconducting bulks. Other circuit model based methods were developed by Morandi in 2014 and Nugteren et al. in 2015 [35][36]. In 2001, Coombs et al. proposed the field-screened method for fast computation of the critical state in type-II superconductors [37]. This approach forced the current density in the element with the maximum vector potential to the critical current density $J_c$ after every iteration step until the external field is screened from the interior. Further studies based on this method were explored by Ruiz-Alonso et al. [38][39]. In 2012, Vesstgarden proposed the fast fourier transform (FFT) based approximation method for modelling the electrodynamics in superconducting thin films [40]. This method has then been utilized mainly for simulating the thermal instabilities and flux avalanches in superconducting thin films [41][42]. In 2015, Shen et al. showed the superiority of the FFT-method over fourier transform and Biot-Savart approaches in calculating the magnetic field from an array of fully magnetized bulks [43], and in 2018, Prigozhin et al. further developed the FFT method and solved 3D magnetization problems related to bulk superconductors [44]. In 2005, Gu et al. proposed the resistivity-adaptive algorithm (RAA) method to calculate the AC loss in HTS tapes [45]. The key concept of RAA is to perform a number of iteration steps to find a resistivity matrix to fulfill the Bean model or the flux creep model. Further studies based on the RAA method were carried out by Farinon et al. in 2010 and by Gu et al. in 2013 [46][47][48]. In 2019, Zhang et al. proposed the direct iteration method to calculate the critical state magnetization currents in a ReBCO tape stack with a large number of layers [49]. The penetration current was updated after each iteration step with a field and angle dependent $J_c(B, \theta)$. In 2020, the same authors extended the iteration method to compute the magnetization of bulk superconductors with both the critical state model and the ***E-J*** power law, for both ascending and descending magnetization stages [50]. In 2020, Zhang et al. proposed a fast and efficient backward computation method to calculate the critical state currents in a 2D periodical bulk HTS undulator [51]. The key concept of the backward calculation is to relax inwards the induced large surface currents gradually, obeying Maxwell's equations and the critical state model.

Apart from the aforementioned numerical methods, FEM is currently the most popular tool for HTS modelling. Such models generally utilize commercial software to implement Maxwell's equations and the ***E-J*** power law or other constitutive laws for HTS modelling [52][53][54][55]. The general forms of the implemented Maxwell's equations can be classified as the ***A-V*** formulation [56][57][58][59][60], the ***T***-Ω formulation [58][61][62][63][64], the ***H***-formulation [65][66][67][68][69][70], the mixed ***T-A*** formulation [71][72][73], the mixed ***H-A*** formulation [74][75][76], and the mixed ***H***-$\varphi$ formulation [77][78][79]. The approximate critical state solution is simulated by setting a large enough *n*-value in the ***E-J*** power law. In 2007, Campbell proposed a new method based on the force-displacement curve of the flux lines to determine the critical state in superconductors [53]. The key concept was to define a flux flow resistivity such that the relevant power law is 1/*n* rather than *n* to speed up the computation and obtain stable solutions. Other similar approaches were later employed by Gömöry et al. in 2012 and Grilli et al. in 2020 [54][55].

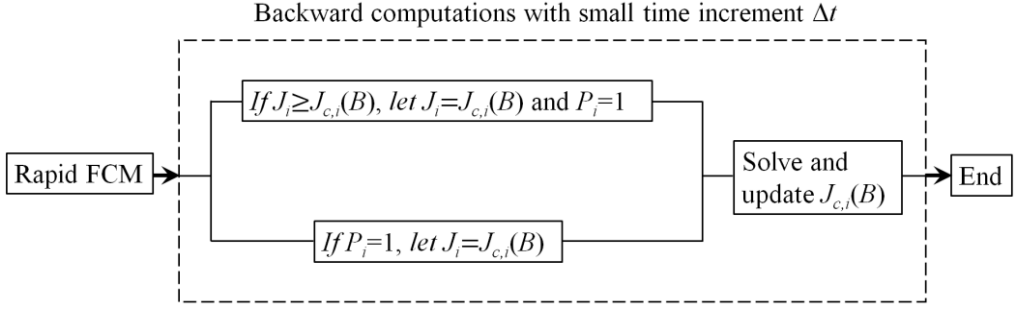

Figure 1. Algorithm for the backward computation method for computing the critical state in a field-cooled magnetized bulk superconductor. *J* refers to the magnetization current density, *i* refers to the element number, and *P* refers to the penetration sign.

To conclude, it has become a popular trend in modelling type-II superconductors to use FEM in the 21$^{st}$ century. However, other numerical methods or eventually combined as well with FEM have also shown excellent performance in modelling the critical state [37][40][45][49][51]. So far, the numerical analysis of magnetization currents in the BHTSU has been carried out with the 3D ***H***-formulation [80][81], the MEM method [82], the RAA-combined ***T***-*Ω* formulation [45][83], and the recently proposed backward computation method [51]. In particular, the backward computation method has shown excellent performance in simulating the magnetization currents in a periodical BHTSU model with 1.8 million DOFs within 1.4 hours [51]. This paper extends the ***A***-*V* formulation-based backward computation method to compute the critical state currents in a ten-period BHTSU in 3D. Both the magnetization currents and the associated undulator field are validated against the well-known ***H***-formulation and the mixed ***H***-*φ* formulation method, as well as with experimental results. Finally, we use the fastest and most efficient ***A***-*V* formulation-based backward computation method to optimize the bulk sizes to minimize the integrals of the undulator field along the beam-axis.

## 2. Finite element method modelling frameworks

2.1 Theory of the ***A***-*V* formulation-based backward computation and its extension to 3D modelling

Large eddy currents can be induced on the surface of a FC-magnetized superconducting bulk when assuming the magnetic flux pinning force is infinitely large. In such a case, the superconducting bulk acts as a permanent magnet, having a uniformly magnetized internal field equal to the initially applied value. However, in reality this situation can never be realized since the flux pinning force in a type-II superconductor is always limited to a finite value. Assuming one superconducting bulk is FC-magnetized slowly under isothermal conditions, the eddy currents will gradually penetrate inwards inside the bulk following a quasi-static critical state model. In the end, the bulk superconductor is magnetized as much as possible with the minimum electro-magnetic entropy production [14], but without generating higher magnetic field than the initially applied [37]. This is indeed the key theory of the backward computation method recently proposed for critical state modelling of type-II superconductors [51]. During the backward iterations, the large surface currents induced by rapid FC magnetization relax inwards step-by-step obeying Maxwell's equations and the critical state model with isothermal assumptions, as shown in the schematic in Figure 1. The relaxation stops when no more superconductor elements can be penetrated. Any additional penetration trying to further minimize the electro-magnetic entropy production will result in the field-screened phenomenon [37].

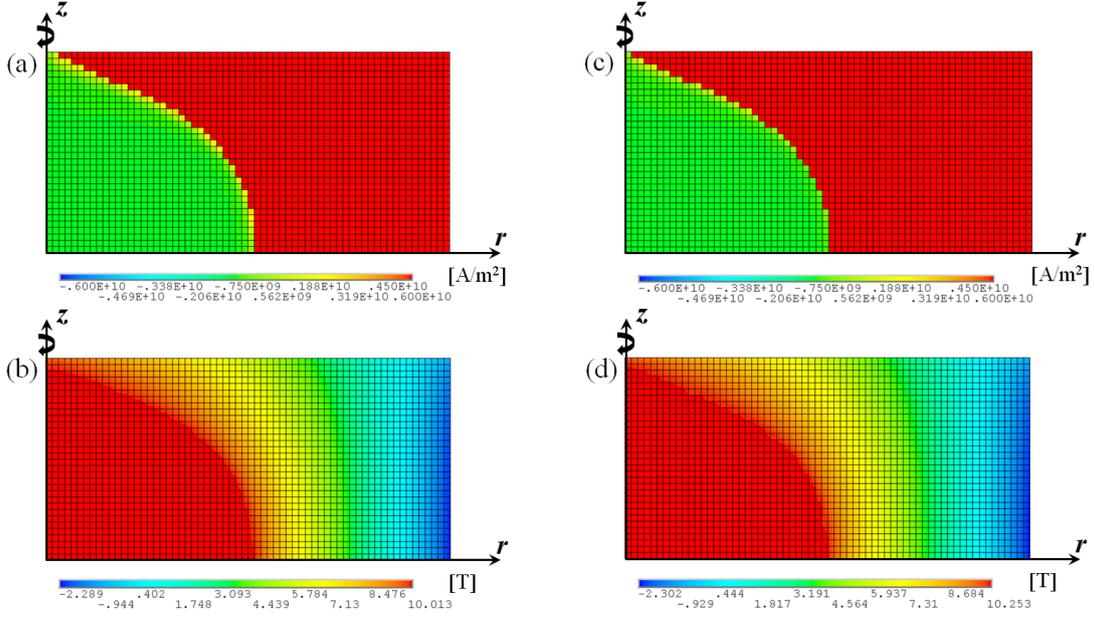

Figure 2. The critical state solution results for (a) magnetization current density $J$ and (b) magnetic field $B_z$ inside a 2D axisymmetric half-bulk model after FC magnetization with an applied field $B_z$ of 10 T. The imaginary solution results for (c) magnetization current density $J$ and (d) magnetic field $B_z$ after introducing an additional penetration layer. The critical current density $J_c$ is assumed to be a constant 6 x $10^9$ A/m$^2$.

The algorithm of the backward computation method can be implemented in any programmable FEM software with different forms of Maxwell's equations. Here we employ the popular FEM software ANSYS using its default *A-V* formulation method, as described by Eqs. (1)-(4).

$$\nabla \times A = B \quad (1)$$

$$\nabla \times B = \mu J \quad (2)$$

$$J = -\frac{1}{\rho}\left(\frac{\partial A}{\partial t} + \nabla V\right) \quad (3)$$

$$\nabla \times \left(\frac{1}{\mu}\nabla \times A\right) = -\frac{1}{\rho}\left(\frac{\partial A}{\partial t} + \nabla V\right) \quad (4)$$

Figures 2(a)-(b) show the critical state solutions of the magnetization current density $J$ and magnetic field $B_z$ inside a 2D axisymmetric half-bulk model after FC magnetization with an applied field $B_z$ of 10 T. It can be observed that the penetrated elements carry a constant current density of 6 x $10^9$ A/m$^2$, obeying the Bean model. The magnetization current density $J$ is not fully zero in the rest of the superconductor elements, especially those nearby the penetration front. This phenomenon, in fact, is quite common in solutions based on FEM and Maxwell's equations, other than numerical methods based on static magnetic field analysis (Biot-Savart law) and iterations of the magnetization currents [22][27][37]. It can be observed that the field component $B_z$ inside the non-penetrated superconductor elements remain around 10 T. This demonstrates well that the backward computation method can maintain the initially applied field in the bulk superconductors as desired. To check whether the critical state solution is comparable to the MEMEP method [14], we have re-calculated the bulk superconductor after artificially introducing an additional penetration layer (one more penetration element in the *x*-direction), as shown in Figure 2(c). Figure 2(d) shows the field component $B_z$ inside the bulk superconductor. It can be observed that the averaged $B_z$ in the unpenetrated

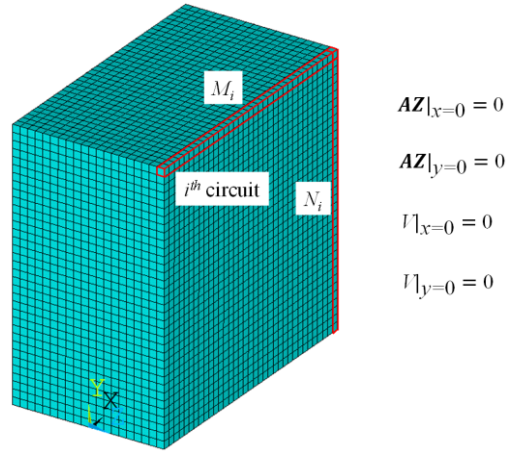

Figure 3. Schematic of a one-quarter cuboid bulk HTS model after element meshing. A uniform magnetic field is applied along the *z*-axis (air subdomain omitted); the nodal voltages at $x = 0$ and $y = 0$ are set to zero to constrain the eddy current simulation; the edge flux at $x = 0$ and $y = 0$ are set to zero to generate flux parallel boundary conditions. $M_i$ refers to the number of elements along the *x*-direction in the $i^{th}$ circuit; $N_i$ refers to the number of elements along the *y*-direction in the $i^{th}$ circuit; the $i^{th}$ circuit has "$M_i + N_i + 1$" elements in total.

superconductor elements is ~10.20 T, larger than the initial 10 T, which is impossible as stated by the field-screened method [37]. To conclude, we can draw conclusion that no more superconductor elements can be further penetrated and the simulation result obtained by the *A-V* formulation-based backward computation method agrees with the MEMEP method.

Studies on the critical state in a superconducting cubic bulk were recently carried out by Pardo et al. using a MEMEP 3D variational method and compared with the 3D ***H***-formulation method [14][84][85]. An obvious off-plane (*x-y*) bending effect for the magnetization currents is observed where far from the mid-plane, since the high induction field at the diagonal is compensated by a $J_z$ component which has opposite signs at the diagonal. However, the assumption of square current loops (i.e., the sand-pile model) can still provide a good approximation of the magnetic moment inside a cuboid bulk HTS [86]. In fact, the current bending effects disappear and the current follows in square loops when the applied magnetic field is above the full penetration field. In this section, we will extend the *A-V* formulation-based backward computation method to calculate the critical state magnetization currents in a 3D cuboid bulk HTS based on the sand-pile model assumption.

Figure 3 shows the schematic of one quarter cuboid bulk HTS model (7 mm x 7 mm x 4 mm, air subdomain omitted) meshed with 3D edge-based finite elements (Solid236 in ANSYS 2020R1 Academic). The edge-flux (***AZ***) DOF is the line integral of the magnetic vector potential ***A*** along the element edge. The edge element-based *A-V* formulation uses tree gauging (GAUGE) to produce a unique solution. The edge-flux ***AZ*** at $x = 0$ and $y = 0$ are set to zero to generate flux parallel boundary conditions and the nodal voltages *V* at $x = 0$ and $y = 0$ are set to zero to constrain the eddy currents. The key concepts of the backward computation are as follows: a) large surface currents are firstly induced in the bulk after rapid FC magnetization from the applied field $B_z$; b) the surface currents relax inwards step-by-step based on the sand-pile model assumption, obeying Maxwell's equations and the critical state model. The constraint of square current loops can be applied by setting the resistivity $\rho$ along the current path to zero (infinitesimal) and the resistivity in the other two orthogonal directions to infinity, as expressed in Eqs. (5)-(7).

$$\rho_x = \begin{cases} 0 & if\ y \geq x \\ \infty & if\ y < x \end{cases} \tag{5}$$

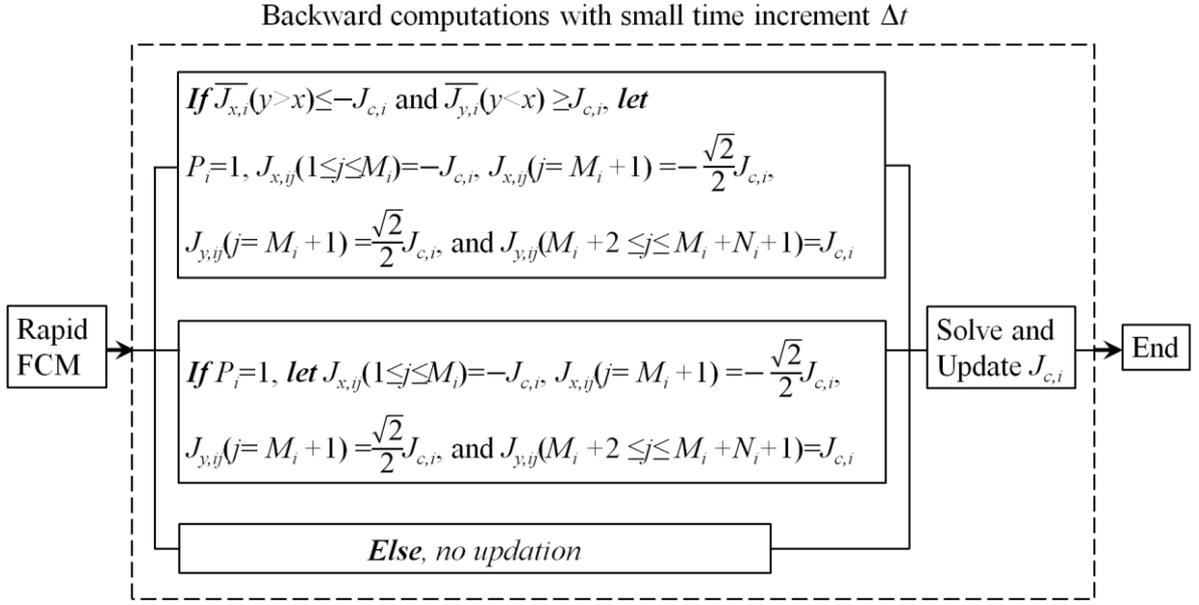

Figure 4. Algorithm for the backward computation method for computing the critical state in a cuboid bulk HTS after FC magnetization from 10 T ($B_z$). $\overline{J_{x,i}}(y>x)$ refers to the averaged x-component of $J_T$ in the top elements; $\overline{J_{y,i}}(y<x)$ refers to the averaged y-component of $J_T$ in the right elements; $J_{x,ij}$ refers to the x-component of $J_T$ for the $j^{th}$ element in the $i^{th}$ circuit; $J_{y,ij}$ refers to the y-component of $J_T$ for the $j^{th}$ element in the $i^{th}$ circuit; $J_{c,i}$ refers to the averaged critical current density in the $i^{th}$ circuit; $P_i$ refers to the penetration sign of the $i^{th}$ circuit.

$$\rho_y = \begin{cases} \infty & \text{if } y>x \\ 0 & \text{if } y \leq x \end{cases} \quad (6)$$

$$\rho_z = \infty \quad (7)$$

Taking the $i^{th}$ circuit in Figure 3 as an example, the resistivity is set to infinity in both the y- and z-directions and zero in x-direction for elements whose center coordinates are "$y > x$"; the resistivity is set to infinity in both the x- and z-directions and zero in the y-direction for elements whose center coordinates are "$y < x$"; the resistivity is set to infinity in the z-direction and zero in both the x- and y-directions for the corner element whose center coordinate is "$y = x$". The critical current density $J_{c,i}$ in the $i^{th}$ circuit is updated after every backward calculation and its value equals the averaged $J_c(B)$, as expressed by Eq. (8). The averaged current densities of $\overline{J_{x,i}}(y>x)$ and $\overline{J_{y,i}}(y<x)$ are calculated for the top elements ($M_i$) and the right elements ($N_i$), respectively, according to Eqs. (9)-(10).

$$J_{c,i} = \frac{1}{M_i + N_i + 1} \sum_{j=1}^{M_i+N_i+1} J_{c,ij}(B) \quad (8)$$

$$\overline{J_{x,i}}(y>x) = \frac{1}{M_i} \sum_{j=1}^{M_i} J_{x,ij} \quad (9)$$

$$\overline{J_{y,i}}(y<x) = \frac{1}{N_i} \sum_{j=M_i+2}^{M_i+N_i+1} J_{y,ij} \quad (10)$$

As shown in the schematic of the numerical algorithm in Figure 4, when $\overline{J_{x,i}}(y>x)$ is lower than $-J_{c,i}$ and $\overline{J_{y,i}}(y<x)$ is greater than $J_{c,i}$, the penetration sign $P_i$ of the $i^{th}$ circuit is set to 1 and the absolute value of the magnetization current density in the $i^{th}$ circuit is forced to $J_{c,i}$; when the value of the penetration sign $P_i$ is 1, the magnetization current density in the $i^{th}$ circuit is updated with the latest $J_{c,i}$. The current direction in the corner element is assumed to be 45 degrees against the *x*- or *y*-direction. The relaxation stops when no more current loops can be penetrated. Assuming the HTS bulk is FC magnetized at 10 K, its critical current density $J_c(B)$ is expressed by the following equation

$$J_c(B) = J_{c1}\exp\left(-\frac{B}{B_L}\right) + J_{c2}\frac{B}{B_{max}}\exp\left[\frac{1}{y}\left(1-\left(\frac{B}{B_{max}}\right)^y\right)\right] \quad (11)$$

where the values of $J_{c1}$, $J_{c2}$, $B_L$, $B_{max}$ and $y$ are 1.0x10$^{10}$ A/m$^2$, 8.8x10$^9$ A/m$^2$, 0.8 T, 4.2 T and 0.8, respectively [51].

## 2.2 3D *H*-formulation-based methods

Two 3D *H*-formulation-based methods are also used to simulate the BHTSU problem: the well-known, traditional *H*-formulation and the recently proposed mixed *H*-φ formulation [77][78][79]. In the 3D *H*-formulation [68], the independent variables are the components of the magnetic field strength, $\mathbf{H} = [H_x, H_y, H_z]$, and the governing equations are derived from Ampere's (Eq. (2)) and Faraday's laws. $\mu$ is assumed to be the permeability of free space, $\mu_0$. The *E-J* power law [87] is used to simulate the nonlinear resistivity of the superconductor, where $\mathbf{E}$ is proportional to $\mathbf{J}^n$ and $n = 100$ is assumed to reasonably approximate the critical state. The *E-J* power law takes into account the assumption made for $J_c$, which may be assumed constant or $J_c(B)$ as described above (Eq. (11)). FC magnetization is simulated as described in [51], by setting an appropriate magnetic field boundary condition on the outer boundaries of the air subdomain such that, for $0 \leq t \leq 100$ s, $\mu_0 H_z(t) = 10 - t/t_{ramp}$, where $t_{ramp} = 100$ s. Thus, the initial condition is $\mu_0 H_z(t = 0\text{ s}) = 10$ T and the magnetic field is ramped linearly down at a rate of 0.1 T/s to $\mu_0 H_z(t = 100\text{ s}) = 0$ T. In the 3D *H*-φ formulation [78], the bulk superconducting subdomains are still modelled using the *H*-formulation, but in the nonconducting, air subdomain, the model solves the magnetic scalar potential φ. This significantly reduces the number of DOFs, since it is a scalar rather than a 3D vector in the *H*-formulation case. In addition, no artificial resistivity is needed in the air subdomain (usually set to 1 Ω·m or similar in the *H*-formulation). In the φ formulation, Ampere's law states that

$$\nabla \times \mathbf{H} = 0 \quad (12)$$

when neglecting displacement currents, and the magnetic scalar potential is defined as

$$\mathbf{H} = -\nabla\varphi \quad (13)$$

In addition, the divergence free condition of **B** results in the governing equation in the air subdomain

$$\nabla \cdot \nabla\varphi = 0 \quad (14)$$

One must also take care to ensure appropriate coupling at the interface between the two formulations: the tangential components of the magnetic field are equated on the *H*-formulation side and the perpendicular components are equated on the φ side as described in [88]. FC magnetization in this case is simulated in the same way as the *H*-formulation for the applied field, and the initial condition is the same for the bulk superconducting subdomains, but in the air subdomain, an initial condition of $\varphi = -10 \cdot z/\mu_0$ (from Eq. (13)) is set.

## 2.3 Large-scale 3D modelling of the BHTSU critical state currents

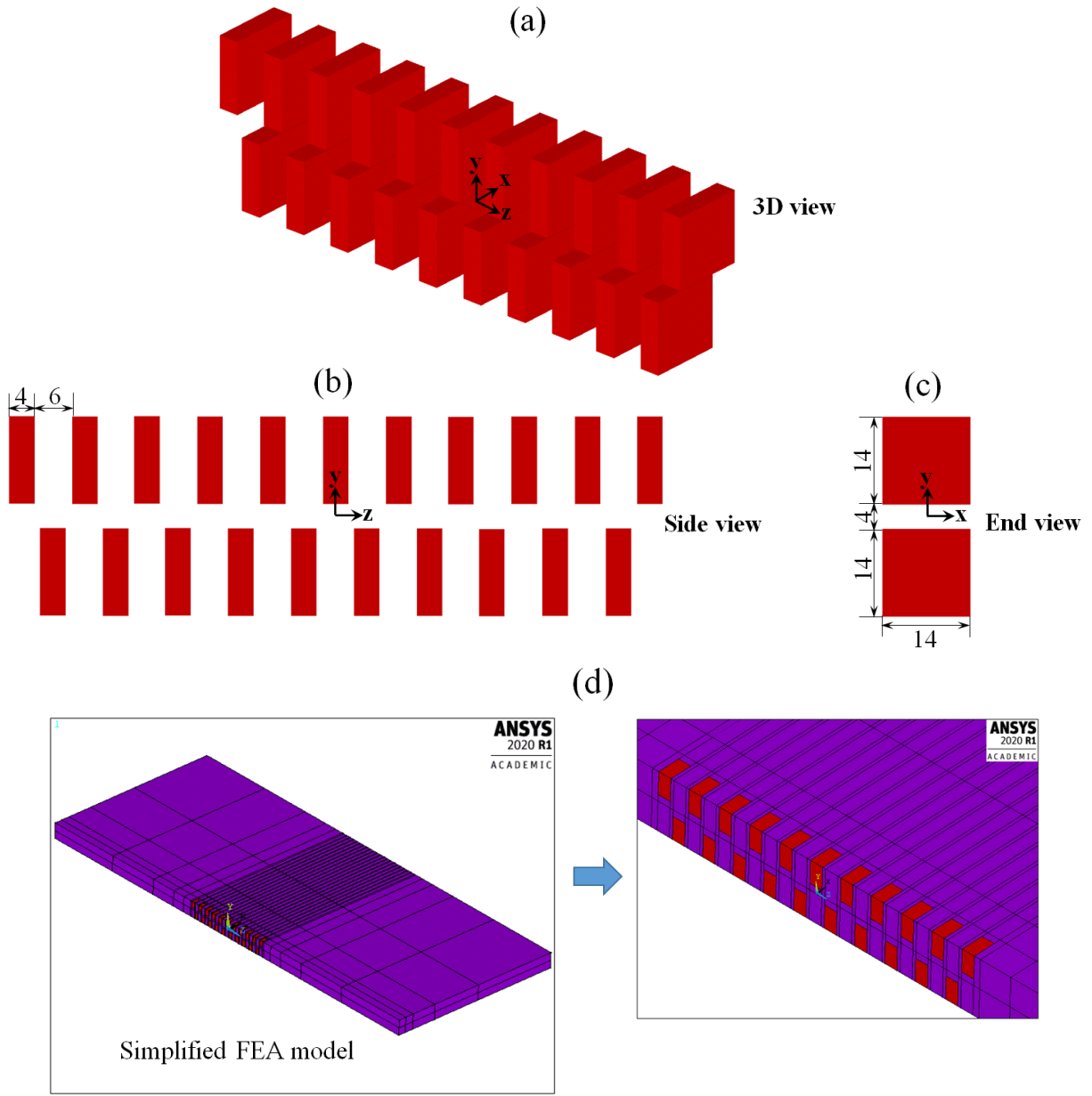

Figure 5. (a) 3D view, (b) side view and (c) end view of the ten-period BHTSU based on staggered-array cuboid bulks, with a period length of 10 mm and a magnetic gap of 4 mm. Each cuboid bulk has the same size: 14 mm x 14 mm x 4 mm. A sinusoidal undulator field $B_y$ is generated along the $z$-axis after FC magnetization from $B_z$ = 10 T. (d) Simplified FEM model of the ten-period BHTSU in ANSYS 2020R1 Academic. The flux parallel boundaries are applied at $x$ = 0 and $y$ = ±9 mm to exploit symmetry; the nodal voltages in the cuboid bulks at $x$ = 0 and $y$ = ±9 mm are set to zero to constrain the eddy current simulation.

As described in [2][4], a sinusoidal undulator field $B_y$ along the electron beam axis ($z$-axis) can be generated in staggered-array superconducting bulks after FC magnetization with a superconducting solenoid. In the models, we approximate this by applying a uniform background magnetizing field. The past prototypes and FEM simulations were all based on staggered-array half-moon-shaped bulk superconductors acting as a compact insertion device [3][4][80][81]. In this section, we study the undulator field generated by staggered-array cuboid bulks, to which the pre-stress can be applied more homogeneously because of their regular shape. Figure 5 (a)-(c) provides different model views of the ten-period BHTSU with a period length of 10 mm and magnetic gap of 4 mm. All the 21 cuboid HTS bulks have the same size: 14 mm x 14 mm x 4 mm. When the electron beams travel along the *z*-axis they are forced to oscillate in the *xz*-plane and generate hard *x*-rays under the action of the Lorentz force associated with the sinusoidal field $B_y$. As the magnetization currents in each cuboid bulk are symmetric based on the assumption of square loops, the ANSYS *A-V* formulation model of the ten-period BHTSU can be simplified as shown in Figure 5(d), with applied flux parallel boundaries at $x = 0$ and $y = \pm 9$ mm. During FC magnetization from 10 T to zero, the temperature inside the ten-period BHTSU is kept constant at 10 K.

The *A-V* formulation-based backward computation method is implemented in ANSYS 2020R1 Academic. To obtain accurate simulation results, the entire FEM model shown in Figure 5(d) is swept meshed with the fine element size of "0.25 mm x 0.25 mm x 0.25 mm" in the HTS bulks and the neighboring air subdomain and with a larger element size in the rest of the air subdomain. All HTS bulks are swept meshed by the edge element Solid236 with the electromagnetic setting (DOFs of *AZ* and *V*) and the air subdomains are swept meshed by the same Solid236 element with the magnetic setting (DOFs of *AZ* only). After element meshing, the whole ANSYS *A-V* formulation model has 0.26 million elements in the HTS bulks, 1.75 million elements in the air subdomains, and 6.46 million DOFs in total; the nodal voltages in HTS bulks are set to zero at $x = 0$ and $y = \pm 9$ mm to constrain the eddy currents.

The *H*- and *H-φ* models were built in COMSOL 5.6: the *H*-formulation is implemented using the Magnetic Field Formulation (mfh) interface and the *φ*-formulation is implemented using the Magnetic Fields, No Current (mfnc) interface. In addition to the general settings outlined in Section 2.2, we make use of the geometric symmetry of the problem to model 1/8$^{th}$ of the problem, as described in [89]. Thus, ¼ of each of the bulks is modelled around the *ab*-plane and ½ of the total undulator is modelled along its length, mirrored across the *xy*-plane along the *z*-axis at $z = 0$. In the *H*-formulation parts, this is achieved using the 'Magnetic Insulation' node (**n** x **E** = 0) for the *yz*-plane at $x = 0$ and *xz*-planes at $y = \pm 9$ mm, and using the 'Perfect Magnetic Conductor' node (**n** x **H** = 0) for the *xy*-plane at $z = 0$, respectively. In the *φ*-formulation, the 'Magnetic Insulation' node is also used, but in this case, **n**·**B** = 0. For the half-symmetry along the undulator length, the 'Zero Magnetic Scalar Potential' node is utilized such that $\varphi = 0$ for the *xy*-plane at $z = 0$. A mapped mesh of element size 0.25 mm x 0.25 mm x 0.25 mm is used in the region $0 \leq x \leq 10$ mm, $-9$ mm $\leq y \leq 9$ mm, $0 < z < 150$ mm and the rest of the air subdomain is meshed as 'Free Tetrahedral' (using COMSOL's predefined 'finer' mesh setting). The model has a total of 0.13 million elements in the HTS bulks and 1.22 million elements in the air subdomain. All elements are linear (first-order) edge (curl) elements. There are then approximately 3.2 million and 1.3 million DOFs for the *H*- and *H-φ* formulations, respectively.

## 3. Results

### 3.1 Comparison of simulation results

Figures 6(a)-(f) summarizes the simulation results of the trapped magnetic field $B_z$ and the magnetization current density $J_s$ in the ten-period BHTSU obtained using the *A-V*, *H*- and *H-φ* formulation methods. The peak $B_z$ in the bulk is 10.0 T, 10.5 T and 10.5 T in the *A-V*, *H* and *H-φ* formulation models, respectively; the peak $J_s$ in the bulk is $8.86 \times 10^9$ A/m$^2$, $8.9 \times 10^9$ A/m$^2$ and $8.9 \times 10^9$ A/m$^2$ in the *A-V*, *H* and *H-φ* formulation models, respectively. As observed in the periodical 2D infinitely long models of the undulator presented in [51], a small flux creep effect exists for the *H*-formulation models, due to the finite (but high) *n* value used in the *E-J* power law and the finite (but slow) ramp-down rate (0.1 T/s).

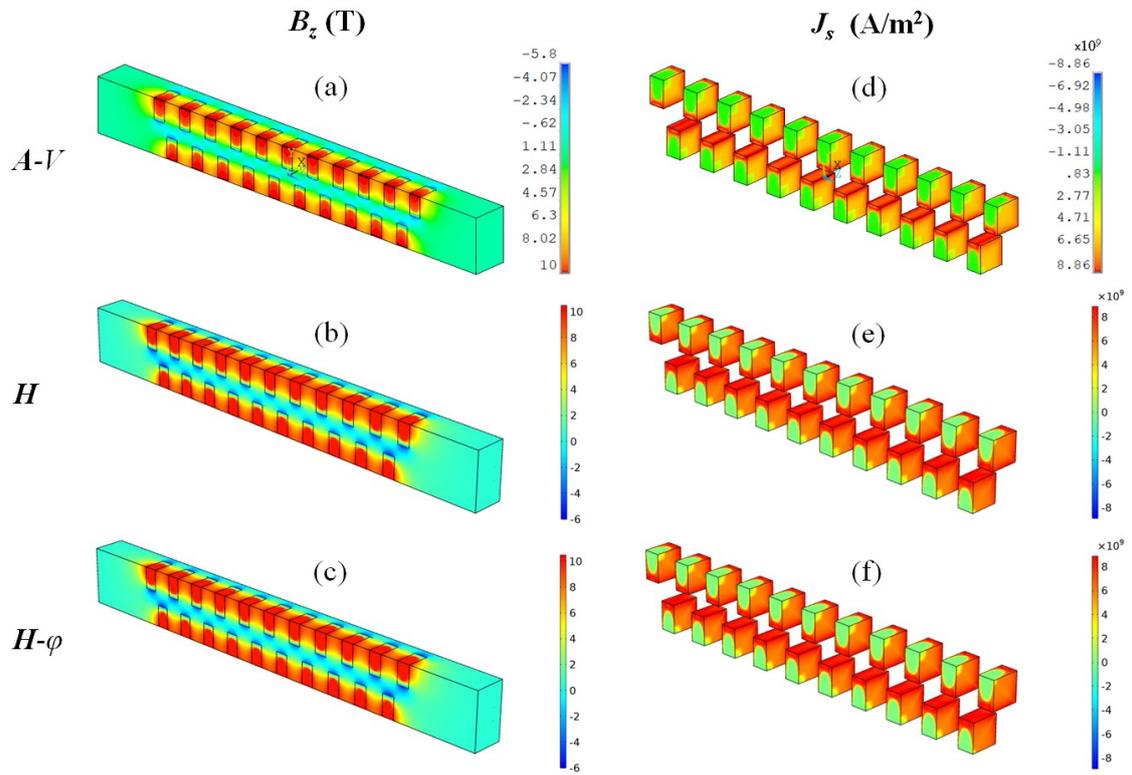

Figure 6. Magnetic field component $B_z$ in the BHTSU obtained using the (a) *A-V*, (b) *H* and (c) *H-φ* formulations. Magnitude of the magnetization current density $J_s$ in the BHTSU obtained using the (d) *A-V*, (e) *H* and (f) *H-φ* formulations. The critical current density $J_c(B)$ assumption for the HTS bulks at 10 K is determined by Eq. (11).

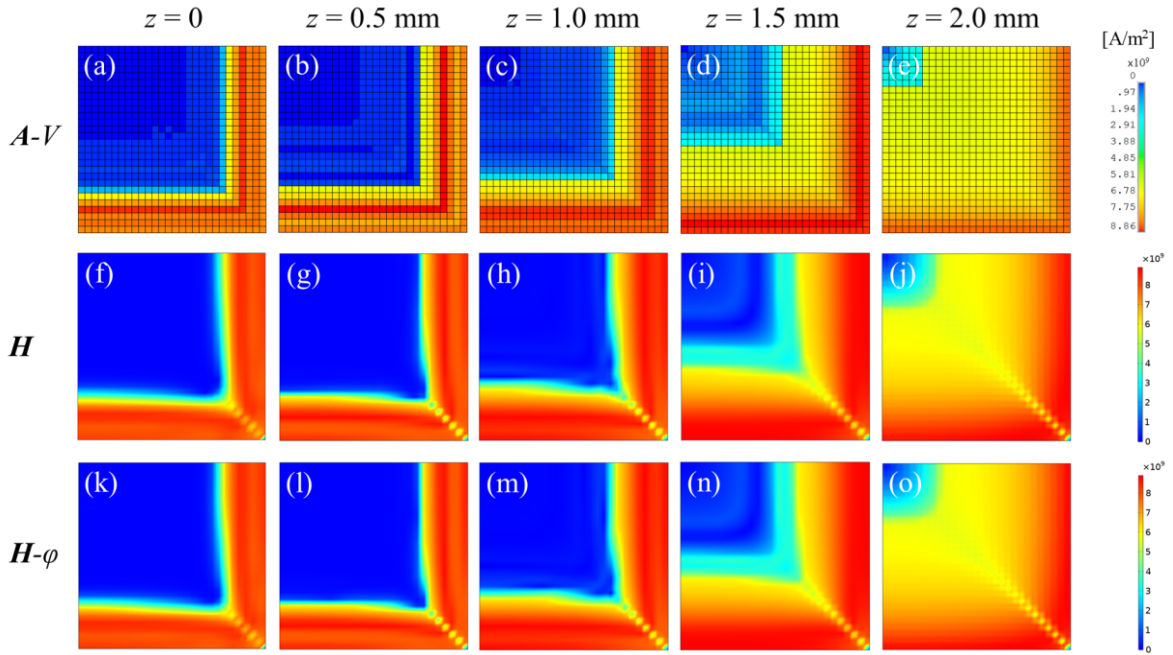

Figure 7. Magnitude of the magnetization current density $J_s$ in the central HTS bulk in the *xy*-plane, solved using the *A-V*, *H* and *H-φ* formulations. "$z = 0$" refers to the mid-plane of the HTS bulk; "$z = 2$ mm" refers to the outer surface of the HTS bulk.

Figure 7 compares the magnitude of the magnetization current density $J_s$ in the central bulk superconductor in the *x-y* plane for each of the models for various positions along the *z*-axis between $z = 0$ (mid-plane of the bulk) and $z = 2$ mm (outer surface of the bulk). While the current density distributions are much the same, there is a subtle but noteworthy difference: the finite geometry of the bulks – in particular, the finite (thin) thickness of the bulks – results in curvature (referred to earlier as current 'bending') of the current density distribution [90][91]. The current loops are necessarily rigid in the *A-V* model due to the constraint of square current loops as assumed for the sand-pile model, but there is no such constraint for the *H*-formulation models.

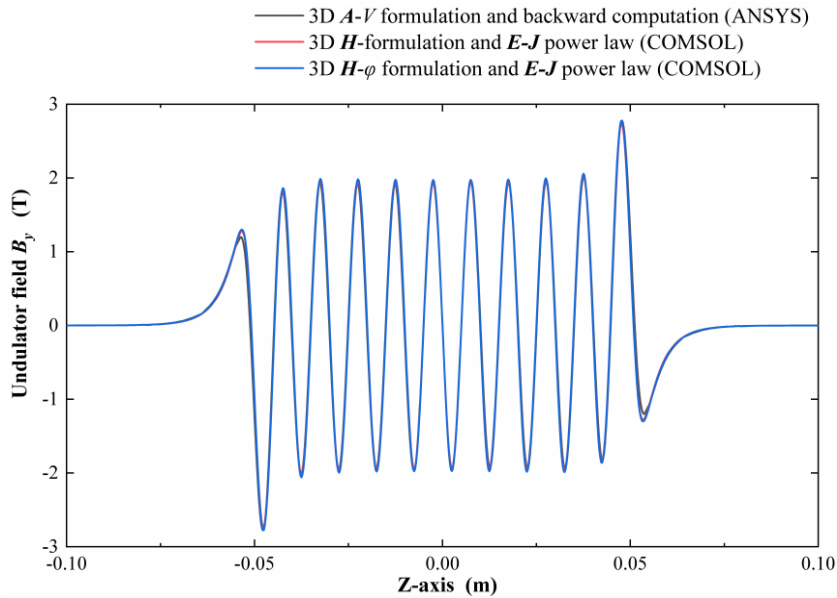

Figure 8. Comparison of the calculated undulator field $B_y$ along the z-axis ($x = y = 0$) obtained using the *A-V*, *H* and *H-φ* formulation models.

Nevertheless, the subtle differences in the trapped magnetic field and current density distributions do not significantly impact the key parameter of interest, the undulator field $B_y$ along the z-axis through the centre of the BHTSU. Figure 8 compares the calculated undulator field $B_y$ along the z-axis obtained using the *A-V*, *H* and *H-φ* formulation models. The three sinusoidal curves show excellent agreement. This indicates that the sand-pile model assumption for the *A-V* formulation-based backward computation method is feasible and that the *H-φ* formulation can provide accurate solutions as a useful and faster alternative to the traditional *H*-formulation.

3.2 Computation time

Table 1. Summary of computation times for the ten-period BHTSU for the three models under investigation

|  | No. of DOFs (million) | Computation time (hour) |
| --- | --- | --- |
| *A-V* | 6.5 | 12.5 |
| *H* | 3.2 | 151 |
| *H-φ* | 1.3 | 23 |

Table 1 summarizes the computation times of the electromagnetic model of the ten-period BHTSU solved by using the *A-V* formulation in ANSYS 2020R1 Academic and the *H*-formulation and the mixed *H-φ* formulation in COMSOL 5.6. All simulations were conducted on a HP-Z8-G4 workstation with 128 GB RAM and two Intel® Xeon® Gold 6128 CPUs @ 3.40

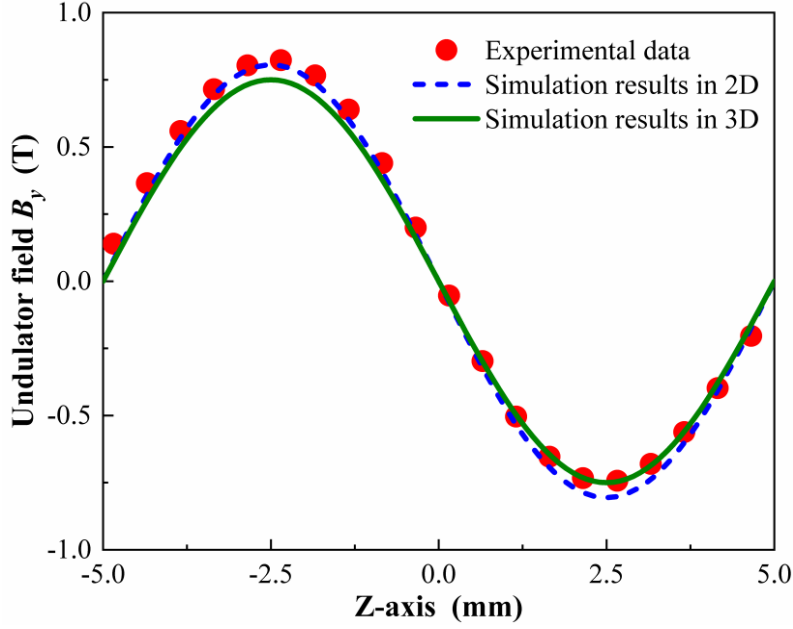

Figure 9. Comparison between the measured undulator field $B_y$ (the central period) after FC magnetization from 6 T at 10 K [4], the simulation results in 2D [51] and the simulation results in 3D obtained by using the *A-V* formulation-based backward computation method. The measurement $B_y$ is taken from the central period of the Gd-Ba-Cu-O bulk undulator prototype.

GHz (12 cores in total). It can be concluded that the *A-V* formulation-based backward computation is the fastest and most efficient numerical method out of the three, solving 6.5 million DOFs within 12.5 h. Due to the reduced number of DOFs, the mixed *H-φ* formulation method has much higher computational efficiency (~60% less DOFs and over six times faster) in comparison to the full *H*-formulation method.

3.3 Comparison with experimental results

A five-period Gd-Ba-Cu-O bulk undulator with a period length of 10 mm and the magnetic gap of 6 mm was tested at the University of Cambridge [4]. As a first attempt to verify the BHTSU concept, FC magnetization experiments were conducted with a conservative background solenoid field of 6 T, rather than the simulated 10 T. Figure 9 shows the calculated undulator field $B_y$ along the *z*-axis of the central period of the 3D BHTSU model (10 mm period length, 6 mm magnetic gap) after FC magnetization from 6 T. The results are compared with the simulation results from a 2D periodical undulator model and the measurement results from the central period of the Gd-Ba-Cu-O bulk undulator prototype [4][51], showing excellent agreement with each other.

## 4. Optimal design of the BHTSU

In synchrotron radiation light sources or *x*-ray free electron lasers, the electron beams are expected to follow their original direction of motion after experiencing the field of an undulator, without neither an offset nor an angle with respect to the orbit defined by the accelerator magnets. It is therefore essential to minimize the first and second integrals of the undulator field $B_y$ along the *z*-axis according to Eqs. (15)-(16).

$$IB_y(z) = \int_{-\infty}^{z} B_y(z')\,dz' \tag{15}$$

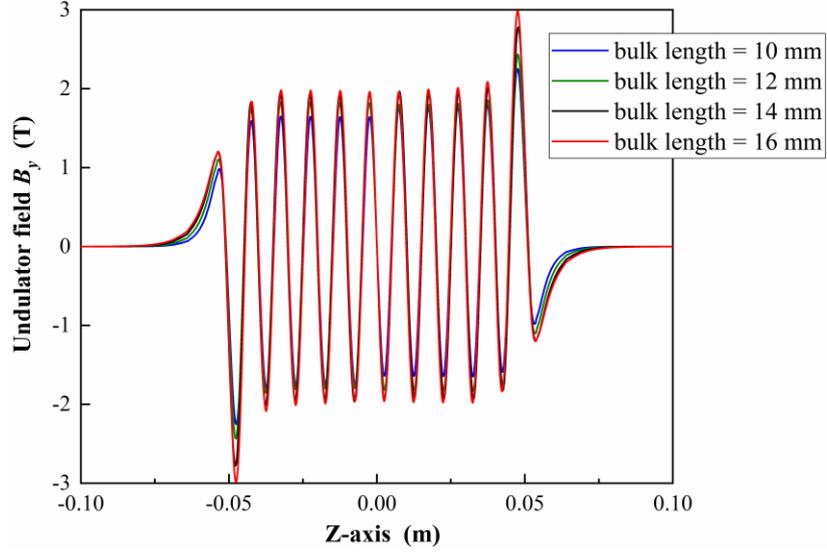

Figure 10. Relation between the undulator field $B_y$ along z-axis and the main bulk sizes. The thickness of the bulks (in the z-direction) is fixed at 4 mm.

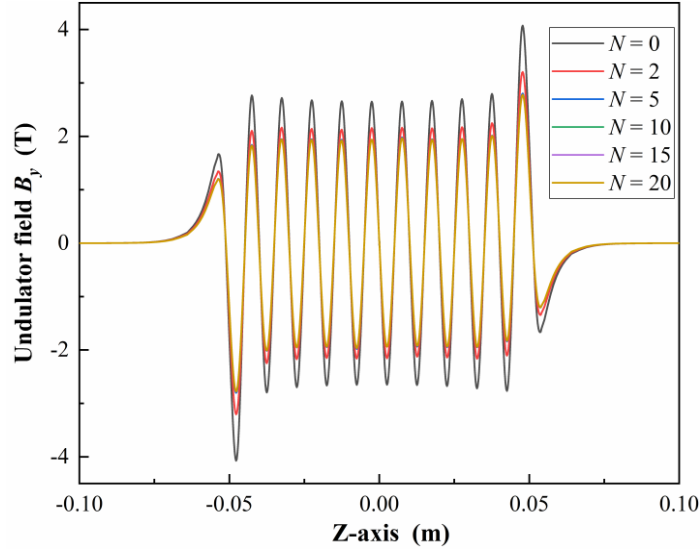

Figure 11. The undulator field $B_y$ after $N$ backward iterations.

$$IIB_y(z) = \int_{-\infty}^{z} IB_y(z')\,dz' \quad (16)$$

The influence of the bulk sizes on the undulator field $B_y$ was investigated by running a series of 3D BHTSU models. As compared in Figure 10, the amplitude of $B_y$ increases as the length of the bulks' sides increases from 10 mm to 14 mm. Further increasing the bulk length does not appreciably increase the undulator field $B_y$. Therefore, the size of the main superconducting cuboid bulks is set to "14 mm x 14 mm x 4 mm". The integral values ($IB_y$ and $IIB_y$) of the undulator field $B_y$ are minimized by optimizing the sizes of the end bulks in the ten-period BHTSU and the associated magnetic gaps. First attempts are made by merely optimizing the sizes of the two end bulks on the top, but a smooth transition of the undulator field $B_y$ cannot be obtained since two field peaks at the ends always exist. Further studies indicate that slightly reducing the sizes of the two end bulks at the bottom can eliminate the two field peaks at the ends. Finally, the size of the two end bulks at the bottom is set to "13 mm x 13 mm x 4 mm"; the integrals of the undulator field $B_y$ are minimized by optimizing the sizes of the two end bulks on the top. The optimization process can be sped up because the undulator field $B_y$ along z-axis often becomes stable after only ~5 backward iterations, as shown in Figure 11. The additional backward iterations help to smooth the magnetization currents near the penetration boundary within the HTS bulks, but this results in a negligible change in the undulator field $B_y$, the key parameter. This allows for a quicker evaluation of $B_y$ along the z-axis and the overall spent time on optimizing the integrals of $B_y$ can be shortened.

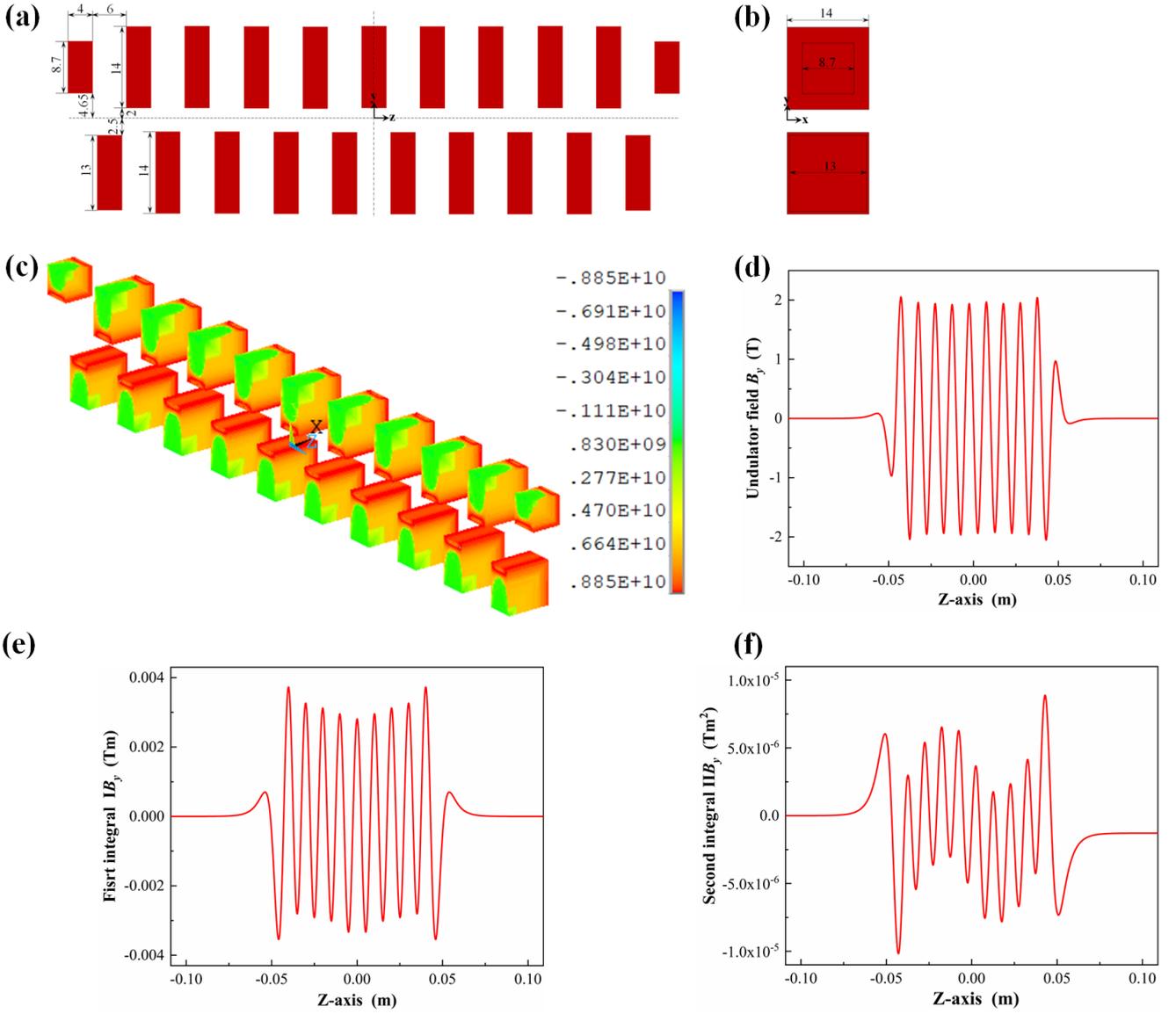

Figure 12. (a) Side view of the optimized BHTSU; (b) end view of the optimized BHTSU; (c) vector sum of the magnetization current density in the simplified BHTSU model; (d) undulator field $B_y$ along the z-axis; (e) first integral of the undulator field $IB_y$ along the z-axis; (f) second integral of the undulator field $IIB_y$ along the z-axis. The size of the main HTS bulks is "14 mm x 14 mm x 4 mm"; the size of the two end HTS bulks on the top is "8.7 mm x 8.7 mm x 4 mm"; the size of the two end HTS bulks at the bottom is "13 mm x 13 mm x 4 mm".

Figures 12(a)-(b) show the side and end view of the optimal design of the ten-period BHTSU. The size of the main HTS bulks is "14 mm x 14 mm x 4 mm"; the size of the two end bulks on the top is optimized to "8.7 mm x 8.7 mm x 4 mm"; and the size of the two end bulks at the bottom is optimized to "13 mm x 13 mm x 4 mm". Figures 12(c)-(d) show the critical state

magnetization current density $J_s$ in the staggered-array HTS bulks and the induced undulator field $B_y$ along the $z$-axis. It should be pointed out that the magnetization currents in the HTS bulks are not fully symmetric with respect to the $xy$-plane because of some small numerical error induced by the iterative analyses. The numerical errors can be corrected by retaining the undulator field $B_y$ in the left-half model and extending the point-symmetric $B_y$ values based on the fact that a symmetric undulator geometry can always generate a point-symmetric undulator field $B_y$, having a zero value of the first field integral $IB_y(-\infty, +\infty)$ along the $z$-axis, as shown in Figure 12(e). Thus the main task is to minimize the second field integral $IIB_y(-\infty, +\infty)$ as much as possible to reduce the offset of the electron beams passing through the HTS undulator. Figure 12(f) plots the integral values of $IIB_y(z)$ along the $z$-axis for the given bulk sizes shown in Figures 12(a)-(b). The optimized $IIB_y(-\infty, +\infty)$ is ~1.2 x $10^{-6}$ Tm$^2$, satisfying the requirements for an insertion device commissioned in the synchrotron radiation light source.

It is worth mentioning that the optimization of the undulator field $B_y$ is based on each of the HTS bulks having the same $J_c$ characteristics (as given by Eq. (11)). The material properties often differ between different HTS bulks, as demonstrated in [4], even within the same batch. In such a case, to improve the accuracy of the models even further, we need to first evaluate the $J_c$ of each HTS bulk by referring to the measured undulator field $B_y$ and then tune the amplitude of $B_y$ and the associated phase errors through moving the HTS bulks up and down [92].

## 5. Conclusion

In this work, the theory of the recently-proposed 2D *A-V* formulation-based backward computation method is extended to calculate the critical state magnetization currents in a ten-period staggered-array BHTSU in 3D. The numerical algorithm is implemented in ANSYS 2020R1 Academic based on the sand-pile model assumption. The simulation results of the undulator field obtained using this ANSYS *A-V* formulation are compared with results obtained using 3D *H*- and mixed *H-φ* formulations implemented in COMSOL, all methods showing excellent agreement with each other as well as with experimental results. In the *A-V* formulation, the ability to set the nature of the DOFs in the Solid236 edge elements makes it possible to perform transient electromagnetic analysis (DOFs: *AZ+V*) in the HTS bulks and static magnetic analysis (DOFs: *AZ*) in the air subdomains. This merit is similar to the mixed *H-φ* formulation, in which the *H*-formulation is adopted in the HTS bulks and the magnetic scalar potential *φ* is solved in the surrounding air subdomain. Compared to the full *H*-formulation, which also calculate the eddy currents in the air subdomains, both the *A-V* and the mixed *H-φ* formulation are shown to have much higher computation efficiency. Solving the ten-period BHTSU, the *A-V* formulation takes 12.5 h for 6.5 million DOFs, faster than the *H-φ* formulation which needs 23 h for 1.3 million DOFs. The faster computation speed of the *A-V* formulation is attributed to the concept of the backward calculation and the absence of a nonlinear resistivity representing the superconducting properties. The fastest and most efficient *A-V* formulation is then adopted to obtain the optimal design of the ten-period BHTSU. After more than twenty iterations of the sizes of the end HTS bulks, the second field integral of the ten-period BHTSU is minimized to 1.2 x $10^{-6}$ Tm$^2$. The numerical algorithm of the 3D backward computation is fast, reasonably straightforward and adaptable to other programmable FEM software packages, allowing it to be readily extended to simulate other large-scale HTS magnetization problems in the future.

**Acknowledgement**

The authors would like to thank Dr. Sebastian Hellmann for his helpful discussions in 3D BHTSU modelling. This work is supported by European Union's Horizon2020 research and innovation program under grant agreement No. 777431 and by the Swiss Accelerator Research and Technology (CHART) program. Dr. Mark Ainslie would like to acknowledge financial support from an Engineering and Physical Sciences Research Council (EPSRC) Early Career Fellowship EP/P020313/1. All data are provided in full in the results section of this paper. The ANSYS APDL codes and COMSOL models in this article will be shared on the website of HTS Modelling Workgroup http://www.htsmodelling.com/.